\documentclass[10pt]{elsart}
\usepackage{epsfig}
\newcommand{\tetaot}{\mbox{$\theta_{13}$}}

\newcommand{\delot}{\mbox{$\Delta_{13}$}}

\newcommand{\nn}{\nonumber}

\newcommand{\be}{\begin{equation}}
\newcommand{\ee}{\end{equation}}
\newcommand{\bea}{\begin{eqnarray}}
\newcommand{\eea}{\end{eqnarray}}

\begin{document}

%
\thispagestyle{empty}
\begin{flushright}
{\tt hep-ph/0108109}\\
{IFT-UAM/CSIC-01-24} \\
{FTUAM 01/16}
\end{flushright}
%
%

\begin{frontmatter}
\begin{center}


\title{Leptonic CP Violation measurement at the neutrino factory}
\vspace{-1cm}



\begin{center}
\author[a]{J. \snm Burguet Castell}\footnote{jordi.burguet-castell@cern.ch}
\author[b]{, O. \snm Mena}\footnote{mena@delta.ft.uam.es}
\address[a]{Dep. de F\'{\i}sica At\'omica y Nuclear and IFIC, 
            Universidad de Valencia, \cny Spain}
\address[b]{Dep. de F\'{\i}sica Te\'orica, 
            Universidad Aut\'onoma de Madrid, 
            \cny Spain}
\end{center}

%
\begin{abstract}
\noindent
\small{In this talk, based on the work \cite{cpviolation}, we refine our previous analysis \cite{golden} of the sensitivity to leptonic CP violation and $\tetaot$ at a neutrino factory in the LMA-MSW scenario, by exploring the full range of these two parameters. We have discovered that there exist, at fixed neutrino energy, $E_\nu$, and baseline, $L$, degenerate solutions. Although the spectral analysis helps in disentangling fake from true solutions, a leftover product of this degeneracy remains for a realistic detector, which we analyse. Furthermore, we take into account the expected uncertainties on the solar and atmospheric oscillation parameters and in the average Earth matter density along the neutrino path.  An
intermediate baseline of $O(3000)$ km is still the best option to
tackle CP violation, although a combination of two baselines turns out
to be very important in resolving degeneracies. This summary is based on a talk given at NUFACT01, May 2001, Tsukuba (Japan).}
\end{abstract}
%
%
\end{center}
\end{frontmatter}
\begin{figure}[ht]
\begin{center}
\epsfig{file=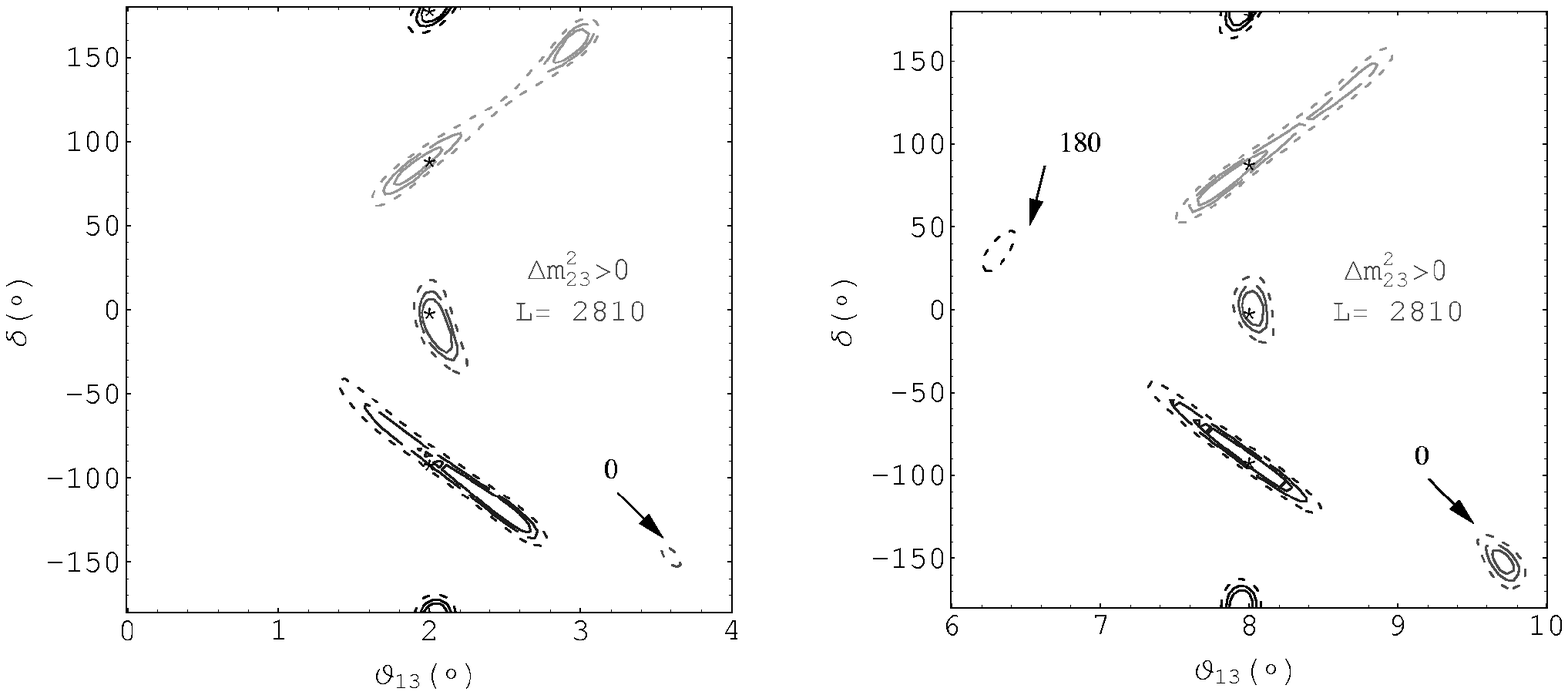, width=10cm,height=5cm} 
\end{center}
\caption{ Simultaneous fits of $\delta$ and $\theta_{13}$ 
at $L = 2810$ km for different central 
values of $\bar{\delta}$ and ${\bar\theta}_{13}= 2^\circ$ (left), $8^\circ$ (right)(atmospheric regime). The value of $\bar{\delta}$ for the degenerate solutions is also indicated.}
\label{golden1}
\end{figure}
\begin{figure}[ht]
\begin{center}
\epsfig{file=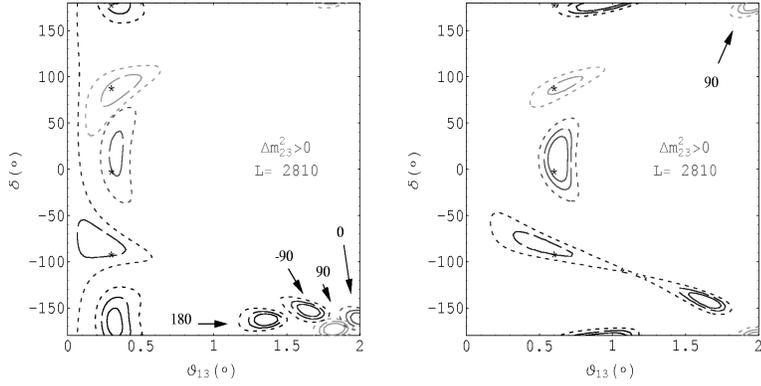, width=10cm,height=5cm} 
\end{center}
\caption{ The same as Fig.~\ref{golden1} for ${\bar \theta}_{13}= 0.3^\circ$ (left), $0.6^\circ$ (right)(solar regime).} 
\label{thpeq2810}
\end{figure}
\begin{figure}[ht]
\begin{center}
\epsfig{file=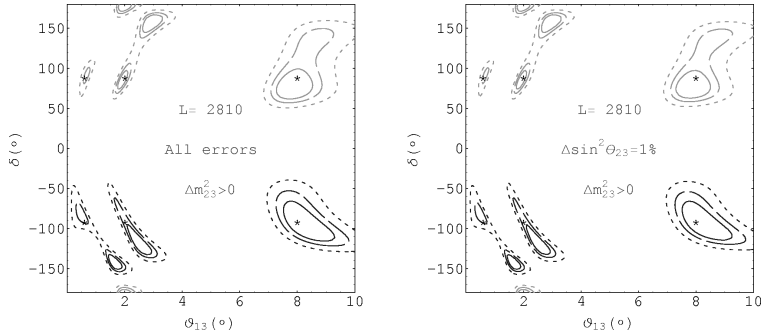, width=10cm} 
\end{center}
\caption{ Fits of $\delta$ and $\theta_{13}$ at $L=2810$ km including 
all the errors  on the remaining parameters (left plot) with $\Delta A/A = 1\%$ and including only the error on $\theta_{23}$ (right plot).}  
\label{allerrors_2810}
\end{figure}
\begin{figure}[ht]
\begin{center}
\epsfig{file=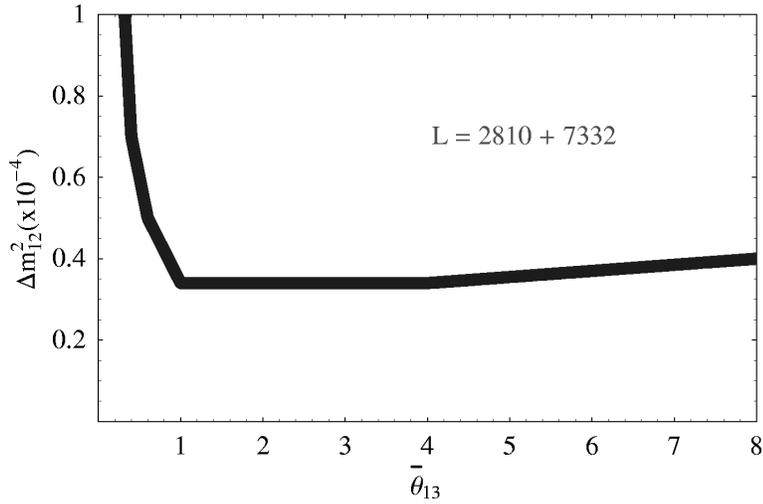, width=10cm}
\end{center}
\caption{ 
Sensitivity reach for CP violation as defined in the text on the plane
$(\Delta m^2_{12}, {\bar \theta}_{13})$ for the combination of baselines
 L = 2810 and 7332 km. All errors are included. }
\label{excl}
\end{figure}
\begin{figure}[ht]
\begin{center}
\epsfig{file=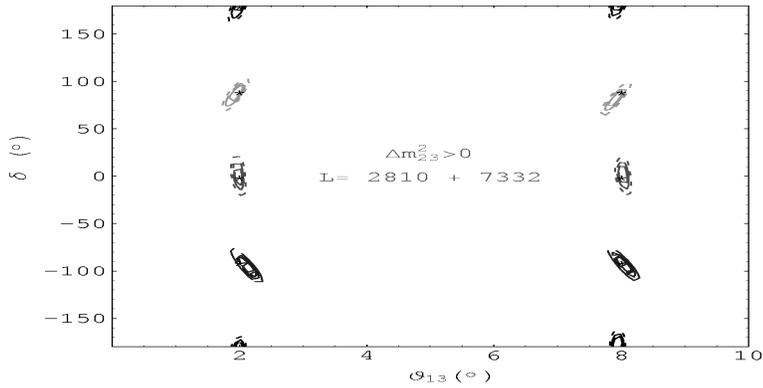,height=5cm,width=10cm}  
\end{center}
\caption{ Fits of $\delta$ and $\theta_{13}$ at the best combination of baselines (atmospheric regime)}
\label{fig:ana2}
\end{figure}
\begin{figure}[ht]
\begin{center}
\epsfig{file=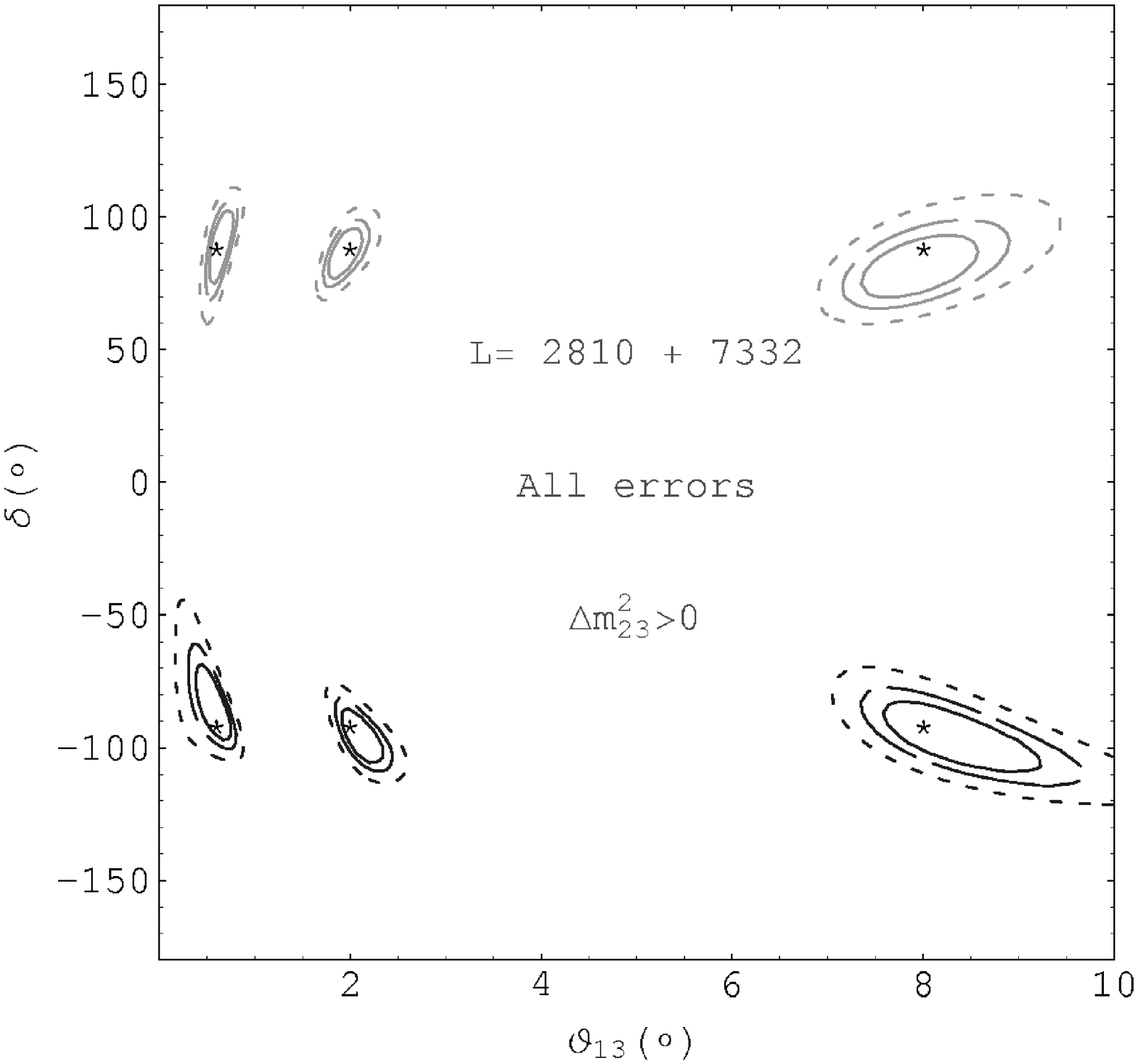, width=10cm,height=5cm} 
\end{center}
\caption{ The same as Fig.~\ref{fig:ana2} with all the errors on the remaining parameters included ($\Delta A/A = 1\%$).}
\label{allerrors_ml}
\end{figure}

%
%
\pagestyle{plain} 
\setcounter{page}{1}
%
%
\newpage
%
\textbf{Sensitivity to CP violation in the LMA-MSW scenario.} 
%
\ The main challenge of the neutrino factory \cite{history,geer} is to measure simultaneously the CP violating phase, $\delta$, \cite{golden,todos1,dgh,todos2} and $\tetaot$ if the solution to the solar neutrino problem is the LMA-MSW scenario \cite{msw,solar}. The best way to determine these two parameters is through the subleading transitions $\nu_e\rightarrow \nu_\mu$ and ${\bar \nu}_e\rightarrow \bar{\nu}_\mu$ by searching for wrong--sign muons \cite{geer,dgh} for both polarities of the beam, i.e. $\mu^+$ and $\mu^-$ respectively. Since there are two small parameters, $\tetaot$ and $\Delta m^{2}_{12}$, (if it is compared with all the relevant energy scales at terrestial distances), a convenient and precise approximation is obtained by expanding to second order in them. Defining $\Delta_{ij} \equiv \frac{\Delta m^2_{ij}}{2 E}$, the result is (details of the calculation can be found in \cite{golden}):
\bea
P_{\nu_ e \nu_\mu ( \bar \nu_e \bar \nu_\mu ) } & = & 
s_{23}^2 \sin^2 2 \tetaot \, \left ( \frac{ \delot }{ \tilde B_\mp } \right )^2
   \, \sin^2 \left( \frac{ \tilde B_\mp \, L}{2} \right) \, + \, 
c_{23}^2 \sin^2 2 \theta_{12} \, \left( \frac{ \Delta_{12} }{A} \right )^2 
   \, \sin^2 \left( \frac{A \, L}{2} \right ) \nn \\
& + & \label{approxprob}
\tilde J \; \frac{ \Delta_{12} }{A} \, \frac{ \delot }{ \tilde B_\mp } 
   \, \sin \left( \frac{ A L}{2}\right) 
   \, \sin \left( \frac{\tilde B_{\mp} L}{2}\right) 
   \, \cos \left( \pm \delta - \frac{ \delot \, L}{2} \right ) \, , 
\label{hastaelmogno}
\eea
where $L$ is the baseline, $\tilde B_\mp \equiv |A \mp \delot|$ and the 
matter parameter, $A$, is given in terms of the average electron 
number density, $n_e(L)$,  as $A \equiv \sqrt{2} \, G_F \, n_e(L)$, 
where the $L$-dependence will be taken from \cite{quigg} and $\tilde J$ is defined as $\tilde J \equiv \cos \theta_{13}\ \sin 2 \theta_{13}\ \sin 2 \theta_{23}\ \sin 2 \theta_{12}$.
In the following we will denote by atmospheric, $P^{atm}_{\nu ( \bar \nu ) }$, solar, $P^{sol}$, and interference term, $P^{inter}_{\nu ( \bar \nu) }$, the three terms in eqs.~(\ref{hastaelmogno}).  
It is easy to show that $|P^{inter}_{\nu(\bar \nu)}| \leq P^{atm}_{\nu ( \bar \nu ) } + P^{sol}$, implying two very different regimes. When $\theta_{13}$ is 
relatively large or $\Delta m^{2}_{12}$ small, the probability is dominated
by the atmospheric term, since $P^{atm}_{\nu (\bar \nu )}\gg P^{sol}$. We will 
refer to this situation as the atmospheric regime. Conversely, when 
$\theta_{13}$ is very small or $\Delta m^{2}_{12}$ large, the solar 
term dominates $P^{sol} \gg P^{atm}_{\nu (\bar \nu )}$. This is the solar
regime. It is essential then to understand whether the correlation between $\delta$ and $\tetaot$ can be resolved in such a way CP violation is measurable in both regimes.  
%
\newline
\textbf{Simultaneous determination of $\delta$ and $\theta_{13}$.}
%
The first question now is if by measuring $P_{\nu_ e \nu_\mu}$ and  $P_{\bar \nu_e \bar \nu_\mu}$, it is posible to  determine unambigously $\delta$ and $\tetaot$ at fixed neutrino energy, $E_\nu$, and baseline, $L$. The answer is no, because at fixed $E_\nu$ and $L$ there exist degenerate solutions for $(\tetaot, \delta)$ which give the same probabilities than some central values chosen by nature $(\bar{\theta}_{13}, \bar{\delta})$. We have performed simultaneous $\chi^2$ fits of the parameters $\delta$ and $\theta_{13}$ for three reference baselines $L=732$ km, $2810$ km and $7332$ km, as well as for various combinations of them. Realistic efficiencies and backgrounds have been included for a 40 Kton magnetized iron detector \cite{lmd}. 
We have considered a muon beam of $50$ GeV (five energy bins of $10$ GeV) providing $10^{21}$ useful $\mu^+$ and $\mu^-$ decays, which is our working setup in the present work, as it was in \cite{golden}. Our detector has very low efficiencies for neutrino energies below $10$ GeV.
All the results that we show correspond to central values of the parameters 
in the LMA-MSW scenario: $\Delta m^2_{12} = 10^{-4}$ eV$^2$,  
$\Delta m_{23}^2 = 3 \times 10^{-3}$ eV$^2$ and
$\theta_{12} = \theta_{23} = 45^\circ$, except in Fig. \ref{excl}, where 
the full range of  $\Delta m^2_{12}$ is considered. Degenerate solutions are clearly seen in our fits even if they include several bins in energy.
%
\newline
\textbf{Atmospheric regime.}
%
\ In Figs.~\ref{golden1} we show the $68.5\%$, $90\%$ and $99\%$ contours resulting from the fits for $L=2810$ km, for four central values 
of ${\bar \delta} =- 90^\circ, 0^\circ, 90^\circ, 180^\circ$ and for 
$\bar{\theta}_{13}= 2^\circ$ (left) and $\bar{\theta}_{13}=8^\circ$ (right). 
The energy dependence of the signals is not significant enough (with our setup) to resolve the expected two-fold degeneracy at the optimal baseline.
%
\newline
\textbf{Solar regime.}
%
\ Figs.~\ref{thpeq2810} depict the analogous to  Figs.~\ref{golden1} for two cases: ${\bar \theta}_{13}= 0.3^\circ$ (left) and ${\bar \theta}_{13}=0.6^\circ$ (right). In the solar regime, at fixed $E_{\nu}$ and $L$, there exist a $\bar{\theta}_{13}=0$ (and  any $\bar\delta$) mimicking solution. Consider for instance the case ${\bar \theta}_{13}= 0.3^\circ$ (left): 
the degenerate images of the four points chosen appear grouped at the right/lower side of the figure. These are the solutions that mimic $\theta_{13} = 0$.
Note that at ${\bar \theta}_{13} = 0.3$, the sensitivity to CP is already lost for ${\bar \delta} = -90^\circ$.  
%
\newline
\textbf{New analysis: inclusion of expected errors on oscillation parameters 
and matter density.}
%
\ Recent analysis of the expected uncertainty in the knowledge of the 
atmospheric parameters at the neutrino factory 
 indicate a $\sim 1\%$ uncertainty in  $\Delta m_{23}^2$ and 
$\sin^2 2 \theta_{23}$ 
\cite{barger}\footnote{ Although these analyses have 
been done for the SMA-MSW solution or assuming that the solar parameters
are known, we will assume that in the LMA-MSW scenario the errors on the 
solar parameters or in the matter term do not change this result.}. 
For the solar parameters in the LMA-MSW regime we include the results of the 
analyses of the Kamland reach \cite{kamland}: $2\%$ error in 
$\Delta m_{12}^2$ and 
$\pm 0.04$ in $\sin^2 2\theta_{12}$, for maximal $\theta_{12}$, both 
at $1\sigma$. For the uncertainty on the matter parameter, $A$, we could not find any estimate in the literature. The dispersion of the different models of the Earth density 
profile \cite{bahcall} indicates an uncertainty of $1$--$2\%$ for 
trajectories which 
do not cross the core, although we consider a range between $1$--$10\%$ for illustration. The most important effects result from the uncertainty in $\theta_{23}$ and in the matter parameter $A$ (once $\Delta m_{12}^2$ and $\sin^2 2\theta_{12}$ are assumed to be known from Kamland as discussed above), with the former affecting mainly the measurement of $\theta_{13}$ and the latter the sensitivity to $\delta$. In
Fig.~\ref{allerrors_2810} (left) we show the results from the fits for ${\bar \delta} = 90^\circ$ and $-90^\circ$ at $L=2810$ km, including all errors 
(with an error in the matter parameter of $1\%$) compared (right) with the situation in which only the error on the atmospheric angle $\theta_{23}$ is included. The two graphics in this figure are almost identical, showing that the dominant error is that of $\theta_{23}$.  
The uncertainty in $A$ is more relevant for the determination of 
$\delta$, although if this error is controlled at the percent level 
the effect is negligible. 
Finally, it is interesting to understand how much of the LMA-MSW range  
can be covered in the discovery of CP violation. This is illustrated in 
Fig.~\ref{excl} with a rough exclusion plot.  For the hypothetical nature values ${\bar \delta} = 90^\circ$  and the best combination of 
baselines, $L=2810+7332$ km, the line corresponds to the minimum value of $\Delta m^2_{12}$ at which 
the $99\%$CL error on the phase reaches $90^\circ$ degrees, and is thus 
indistinguishable from $0^\circ$  or $180^\circ$ (i.e. no CP violation). 
All errors on the parameters have been included. With this definition, there is sensitivity to CP violation for $\tetaot>$ few tenths of degree and 
$\Delta m^2_{12}> 3\times 10^{-4}$ eV$^2$.
%
\newline
\textbf{Conclusions.}
%
\ At the hypothetical time of the neutrino factory, the value of 
the parameters $\tetaot$ and $\delta$ may be still unknown and will have to be 
simultaneously measured.
 A relevant problem unearthed is the generic existence, at a given
(anti)neutrino energy and fixed baseline, 
of a second value of the set ($\tetaot,\delta$) which gives the same 
oscillation probabilities for neutrinos and antineutrinos than the 
true value chosen by nature. These degeneracies can dissappear if there exist a significant energy or baseline dependence, but the former, with our statistics and including realistic efficiencies and backgrounds in our spectral analysis, is not strong enough to resolve degeneracies at an optimal baseline. 
We have performed simultaneous fits for the combination of any two baselines in the atmospheric regime. In Fig.~\ref{fig:ana2} we show the result for the best combination of baselines in the atmospheric regime. While the two-fold  degeneracy dissappear completely in the combination of the larger baseline with the two shorter ones, it does not disappear in the combination of the two shorter baselines. In the solar regime, as in the atmospheric one, the degeneracies are nicely resolved in the combination of the intermediate and long baselines. We have also found that degenerate solutions do not exist increasing our statistics by a factor five at the intermediate baseline with our setup. If realistic detectors with a lower detection threshold are feasible \cite{monolith}, the situation can be easier with just one baseline.
 Furthermore, we have included in the analysis the expected uncertainty on 
the knowledge of the rest of the oscillation parameters ($\sin^2 \theta_{23},\,
\Delta m^2_{23},\,\sin^2\theta_{12},\,\Delta m^2_{12}$) and on the Earth 
electron density. Noticeable changes result from the error on $\theta_{23}$, which affects mainly the uncertainty in $\theta_{13}$, and from the 
uncertainty on the Earth matter profile, which affects mainly the extraction of
$\delta$. In Fig.~\ref{allerrors_ml} we show the results for the best combination of baselines when all errors have been included. The resolution of the
degeneracies discussed before is still achieved, but the contours have become sizeably larger. The overall conclusion is that the optimal distance for studying CP-violation effects with neutrino energies of few dozens of GeV is still of $O(3000)$ km, although the combination of two baselines, one of which being preferably a very long one, is very important in resolving degeneracies. 
%
\linespread{1}

\newpage
%

\begin{thebibliography}{9}
%
\bibitem{cpviolation} J.~Burguet Castell, M.B.~Gavela, J.J.~G\'omez C\'adenas, P.~Hern\'andez and O.~Mena, Nucl.\ Phys.\ B {\bf 608} (2001) 301.
\bibitem{golden} A.~Cervera {\it et al.},  Nucl. Phys. {\bf B579} (2000) 17; Erratum-ibid. {\bf B593} (2001) 731.
\bibitem{history} D.G.~Kosharev, CERN internal report CERN/ISR-DI/74-62 (1974).
\bibitem{geer}  S.~Geer, Phys. Rev. {\bf D57} (1998); 6989; and erratum.
\bibitem{todos1} J.~Arafune, M.~Koike and J.~Sato, 
Phys. Rev. {\bf D56} (1997) 3093. Phys. Lett. B{\bf 345} (1998) 373. H.~Minakata and H.~Nunokawa, Phys. Lett. {\bf B413} (1997) 369; Phys. Rev. {\bf D57} (1998) 4403. S.M.~Bilenky, C.~Giunti and W.~Grimus, Phys.Rev. {\bf D58} (1998) 033001. M.~Tanimoto, Phys. Lett. {\bf B462} (1999) 115. 
\bibitem{dgh} A.~de~R\'ujula, M.~B.~Gavela and P.~Hern\'andez, 
Nucl. Phys. {\bf B547} (1999) 21.
\bibitem{todos2} K.~Dick {\it et al}, Nucl. Phys. {\bf B562} (1999) 299. A.~Donini {\it et al}, Nucl.Phys. {\bf B574} (2000) 23. A.~Romanino, Nucl. Phys. {\bf B574} (2000) 675. G.~Barenboim and F.~Scheck, Phys.Lett. {\bf B475} (2000) 95.
M.~Freund {\it et al}, Nucl. Phys. {\bf B578} (2000) 27. J.~Bernabeu and M.C.~Banuls, Nucl. Phys. Proc. Suppl. {\bf 87} (2000) 315. A.~Bueno, M.~Campanelli and A.~Rubbia, Nucl. Phys. {\bf B589} (2000) 577. V.~Barger, S.~Geer and K.~Whisnant, Phys. Rev. {\bf D61} (2000) 053004. H.~Minakata and H.~Nunokawa, Phys. Lett. {\bf B495}(2000)369. S.J.~Parke and T.J.~Weiler, Phys. Lett. {\bf B501}(2001) 106. P.~Lipari, hep-ph/0102046.
\bibitem{msw} L.~Wolfenstein, Phys. Rev. {\bf D17} (1978) 2369; {\bf D20} (1979) 2634; S.P.~Mikheyev and A.~Smirnov, Sov. J. Nucl. Phys. {\bf 42} (1986) 913.
\bibitem{solar} For a recent global analysis of solar data and atmospheric data
 see for instance M.C.~Gonz\'alez-Garc\'{\i}a {\it et al}, Phys. Rev. {\bf D63} (2001) 033005.
\bibitem{quigg} R.~Gandhi {\it et. al.}, Astropart. Phys. {\bf 5} (1996) 81. 
\bibitem{lmd} A.~Cervera, F.~Dydak and J.J.~G\'omez-Cadenas, Nucl. Instrum. Meth. {\bf A451} (2000) 123.
\bibitem{barger} V.~Barger {\it et. al.}, Phys. Rev. {\bf D62} (2000) 013004;\  M.C.~Gonz\'alez-Garc\'{\i}a, talk at the CERN Nufact 
neutrino oscillation working group, February 2000. 
\bibitem{kamland} V.~Barger, D.~Marfatia and B.P.~Wood, hep-ph/0011251. H.~Murayama and A.~Pierce, hep-ph/0012075. 
\bibitem{bahcall} 
J.N.~Bahcall and P.I.~Krastev, Phys. Rev. {\bf C56} (1997) 2839.
J.N.~Bahcall, private comunication.
\bibitem{monolith}
 M.~Freund, P.~Huber and M.~Lindner, hep-ph/0105071.
\end{thebibliography}
\end{document}